\documentclass{emulateapj}

\submitted{Submitted 30 June 2004; Accepted 30 August 2004 -- To appear in 20 December 2004 ApJ}

\lefthead{Werk, Jangren \& Salzer}
\righthead{Luminous Compact Blue Galaxies in KISS.}

\newcommand{\etal}{{\it et\thinspace al.}\ }

\begin{document} 


\title{Luminous Compact Blue Galaxies in the Local Universe}

\author{Jessica K. Werk, Anna Jangren, and John J. Salzer}
\affil{Astronomy Department, Wesleyan University, Middletown, CT 06459; jessica@astro.wesleyan.edu, anna@astro.wesleyan.edu,
 slaz@astro.wesleyan.edu}

\begin{abstract}

We use the KPNO International Spectroscopic Survey (KISS) for emission-line galaxies to identify and 
describe a sample of local analogues to the luminous compact blue galaxies (LCBGs) that are observed to be
abundant at intermediate and high redshift.  The sample is selected using criteria believed effective at 
isolating true examples of LCBGs: $SB_{e}$(B-band) $< 21.0$ mag arcsec$^{-2}$, $M_{B}< -18.5$ (for 
$H_0= 75$ km s$^{-1}$ Mpc$^{-1}$), and $B-V < 0.6$.  Additionally, all LCBG candidates presented are 
selected to have star-formation as their dominant form of activity.  We examine the properties of our 
LCBGs and compare them to those of other KISS star-forming galaxies of the same absolute magnitude range.
We find that the KISS LCBGs lie on the extreme end of a fairly continuous distribution of ``normal'' 
star-forming galaxies in the plane of surface brightness versus color.  This result differs from the
results of previous studies that show LCBGs at higher-z to be more separate from the ``normal'' (usually non-active) 
galaxies they are compared against.  On average, LCBGs have a higher tendency to emit detectable flux in the radio 
continuum, have higher H$\alpha$ luminosities by a factor of 1.6, indicating strong star-formation activity, 
and have slightly lower than expected metal abundances based on the luminosity-metallicity relation for KISS 
galaxies.  We calculate the volume density of our low-$z$ ($z<0.045$) sample to be  
$5.4\times10^{-4}h_{75}^{3}Mpc^{-3}$, approximately 4 times lower than the volume density of the LCBGs at 
0.4 $<$ z $<$ 0.7 and $\sim$10 times lower than the volume density of the population at 0.7 $<z<$ 1.0.

\end{abstract}


\section{Introduction}
\label{intro}

Common at intermediate and high redshifts, yet rare in the local universe, luminous compact 
blue galaxies (LCBGs) typically have luminosities equal to or greater than that of the Milky
Way but are considerably smaller in size \citep{koo94,phil97,guz97,ostlin,cfrs}. As evidenced by such properties,
they are among the most extreme galaxies known in the universe. Although these
galaxies have been the subject of numerous comprehensive studies in the past decade,
 their evolution and nature, to a large extent, remains in dispute \citep{koo95,phil97,hammer,ostlin2,barton,
pisano,guz03}. At the heart of the issue lies the difficulty of identifying nearby analogues that would
allow a more complete exploration of the mechanisms responsible for their extreme properties at 
high redshifts, as well as their evolutionary paths. 

Characterized by high surface brightnesses, small half-light radii, and vigorous star-formation,
LCBGs were initially identified as stellar
objects from 4-m plate surveys for QSO candidates \citep{koo92,koo94,guz96,guz98}. 
The formal definition of LCBGs is developed by \cite{anna} from  examining regions of six-dimensional parameter space 
(color, luminosity, asymmetry, concentration, size, and surface brightness) occupied
by three classes of compact galaxies chosen to represent the population of luminous, blue, compact
galaxies as a whole. The three classes are:
 (1) Compact Narrow Emission-Line Galaxies (CNELGs, Koo \etal 1994, 1995; Guzm\'{a}n 1996, 1998),
 (2) a subset of galaxies very similar to CNELGs studied by \cite{koo95,phil97} and \cite{guz97} called faint blue galaxies,
 and (3) higher redshift Blue Nucleated Galaxies (BNGs, Schade \etal 1995, 1996). While the classification of an LCBG 
relies on the galaxy's falling within three of five defined regions 
in the aforementioned parameter space, recently groups have adopted three cut-offs that roughly
correspond to the criteria laid out by \cite{anna}: $B-V < 0.5-0.6$, $SB_e < 21-21.5$ mag arcsec$^{-2}$, and $M_B <
-18.5$, with $H_0= 70$ km s$^{-1}$ Mpc$^{-1}$ \citep{garland,pisano}. These selection criteria will be further examined 
in subsequent sections.

With a volume density that drops off substantially from $z=1$ to the present (previously estimated to be a change of roughly a 
factor of ten), LCBGs appear to be the most strongly evolving 
galaxy population \citep{lilly96,phil97,guz97,lilly98,cfrs}. \cite{guz97} estimate that they contribute nearly $40\%$ 
 of the fractional increase in the SFR density between $z=0$ and $z=1$. Furthermore, luminous, compact star-forming
galaxies appear to dominate high-redshift galaxy samples. \cite{low97} have suggested that the Lyman Break 
Galaxies at $z=3$, due to their very compact cores and high surface brightnesses, may be the high-redshift 
counterparts of the intermediate-z LCBGs. \cite{smail98}, in a sub-millimeter galaxy survey,
 found a $z\sim4$ galaxy population that may contribute a substantial fraction of the SFR density 
in the early universe, nearly half of which are very luminous and compact. Evidently, LCBGs play a prominent, if not 
paramount, role 
in the formation and evolution of galaxies in the young universe. 

A drove of unanswered questions hovers around the subject of LCBGs, mainly owing to the lack of 
a representative local sample. One of these questions involves the mechanism(s) responsible for the observed 
intense starbursts in LCBGs. 
 In theory, these triggering mechanisms could be determined
by examining the H$\alpha$ velocity fields and morphologies of local LCBG examples \citep{ostlin}.
 This sort of study relies on the assumption
that one can select a sample of local LCBGs to have the same properties as the more distant LCBGs. \cite{ostlin} use a
sample of six galaxies with $M_{B}$ between -17.5 and -20 selected from local blue compact galaxies (BCGs) that are bright
in H$\alpha$ to infer certain kinematic properties of the higher-$z$ LCBGs. They conclude that, in most cases, 
mergers or strong interactions trigger the intense starbursts of LCBGs. However, one should use caution in 
interpreting these results given the size and properties of their sample. Indeed, a statistically complete sample 
of low-z LCBGs will provide more clues as to the the origin of their activity. 

Another crucial topic in the study of LCBGs has been the question of what sort(s) of galaxy they 
become as they evolve. The determination of reliable mass estimates for the intermediate-$z$ LCBGs
is key to establishing these evolutionary connections to present-day galaxies. However, accurate
stellar mass estimates are difficult to determine and measurements of emission-line widths used to indicate 
the virial mass may underrepresent the gravitational potential for the intermediate-redshift LCBGs 
\citep{guz01,guz03,pisano}.  As a consequence of these difficulties, there is not good agreement
for the mass estimates of LCBGs and there is a corresponding uncertainty as to their proposed evolutionary 
path(s).  \cite{phil97}, \cite{barton}, and \cite{hammer} argue for an evolutionary scenario in which LCBGs 
represent the formation of the bulges of objects that will eventually evolve into massive spiral disks of 
today. Koo \etal(1995), and Guzm\'{a}n \etal (1996, 1997, 1998) maintain that while LCBGs represent a 
heterogeneous class of galaxies, the majority will evolve into local low-mass dwarf elliptical galaxies. 

A possible solution to the mass debate may lie in the HI line widths of local LCBGs. These line widths 
can be used to infer the total dynamical mass.  Accordingly, \cite{pisano} set out to 
measure the HI line widths of a nearby sample of galaxies they believe are 
representative of the more distant LCBGs. They infer the masses of the intermediate-$z$ galaxies from these local examples, 
and find that LCBGs have a variety of dynamical masses, and thus a variety of evolutionary paths. Yet, they recognize
that their local sample may not have been truly representative of the the high-$z$ LCBG population. Of their 11 local galaxies, 
only three meet the criteria proposed by \cite{anna}. In a study of 20 local LCBG candidates selected to more nearly meet the 
the criteria of \cite{anna},
 \cite{garland} found results similar to those of \cite{pisano}. Specifically, they show that while most LCBGs are 
nearly ten times less massive than local galaxies of the same luminosity, some are just as massive. The implication
is that LCBGs are a heterogeneous class of galaxy.

Clearly, defining reliable local samples of LCBGs is central to the process of understanding their nature
and evolutionary paths.  The \cite{pisano} and \cite{garland} samples are recent attempts to construct samples
of nearby LCBGs.  The latter study is based on a sample of LCBGs discovered by \cite{castander} who analyzed
data from the Sloan Digital Sky Survey.  Another recent study by \cite{drinkwater} utilized the complete
spectroscopic study of a 12 deg$^2$ area centered on the Fornax Cluster to identify a sample of 13 LCBGs with 
redshifts less than 0.21.  In the current study, we take a slightly different approach.

Using the KPNO International Spectroscopic Survey (KISS, Salzer \etal 2000), we aim to identify and
describe a statistically complete local sample of star-forming galaxies that best resemble those LCBGs 
observed in the more distant universe.  Our low-z sample will provide a basis for future work aimed at 
resolving the debates over the evolutionary histories and kinematics of LCBGs. Due to the limited resolution of the 
survey images, the classification scheme developed by \cite{anna}, that uses morphological parameters,
cannot be employed. Instead, we apply selection criteria consistent with the parameters of \cite{anna} as a way 
to isolate the LCBGs from other KISS star-forming galaxies. While sharp boundaries in parameter space 
do not always effectively isolate one type of galaxy from another, LCBGs in the color-surface brightness plane occupy 
a well-defined region of parameter-space \citep{anna}. Therefore, we employ criteria in \emph{B}$-$\emph{V} color, in 
surface-brightness, and in luminosity to select a sample of local LCBG candidates.  Still, none of the selection criteria we use
can guarantee that the KISS LCBG candidates, like other putative local samples, will contain \emph{bona fide} LCBGs; they
only insure a similarity. Regardless, we believe
that the sample we define will provide an adequate starting point for future investigation. In $\S$ 2, we describe KISS,
our measurements, and selection criteria. An analysis of the properties of our local LCBGs along with the volume density calculation
appears in $\S$ 3.  Section 4 contains a discussion of the implications our sample has for high-$z$ studies and potential future work 
on these galaxies beyond KISS. In $\S$ 5, we summarize our results. Throughout this paper we adopt $H_0= 75$ km s$^{-1}$ Mpc$^{-1}$.

\section{The Data}

\subsection{KPNO International Spectroscopic Survey}
\label{KISS_sum}

The KPNO International Spectroscopic Survey (KISS) aims to find a quantifiably complete, well-defined
sample of extragalactic emission-line sources \citep{john00}. Although this survey is not the first to 
look for galaxies that display unusual activity, it reaches substantially deeper than other wide-field 
objective-prism surveys,
mainly because it $is$ the first survey of this type to employ a large-format CCD as its detector. The survey data, composed of
both objective-prism
and direct images, were  taken
with the 0.61 m Burrell Schmidt telescope located on Kitt Peak. The objective-prism 
images cover a spectral range of either 4800 - 5500 \AA \ or 6400 - 7200 \AA\, and the direct images are
observed through standard \emph{B} and \emph{V} filters. Our local sample of LCBGs is derived
from the first three KISS survey strips: KISS red list 1 (KR1), described in \cite{john01}; KISS blue list 1 (KB1),
 described in \cite{john02};
and KISS red list 2 (KR2), described in \cite{kiss43}. 
KR1 and KR2 use the H$\alpha$ line for selection 
whereas KB1 detects objects by their [OIII] emission line.
The KR1 portion of KISS finds 1128 ELG candidates in a survey area of 62.2 deg$^2$ (or 18.1 KISS ELGs per square degree); 
the KB1 portion of KISS finds 223 ELG candidates in 116.6 deg$^2$ (or 1.91 KISS ELGs per square degree);
and the KR2 portion of KISS finds 1029 ELG candidates in 65.8 deg$^2$ (or 15.6 KISS ELGs per square degree).
KR1 and KB1 overlap with
part of the Century Survey (Geller \etal 1997; Wegner \etal 2001) while 
 KR2 runs through the center of the Bo\"{o}tes void.

The KISS data are run through several steps of processing using a series
of Image Reduction and Analysis Facility\footnote{IRAF is distributed by the National Optical Astronomy Observatory, 
which is operated by the Association of Universities for Research in Astronomy, Inc., 
(AURA) under cooperative agreement with the National Science Foundation.}(IRAF) scripts:
object detection and inventory, photometry and object classification, astrometry, spectral image
coordinate mapping and background subtraction, spectral extraction and overlap correction, emission
line detection, and spectral parameter measurement. Measurements
of the extracted objective-prism spectra yield estimates of the redshifts, line fluxes, and equivalent widths
of the ELGs. For a complete description of the data processing, see \cite{john00}.
 It should be noted that many objects are unresolved in the
direct images (which have a resolution of $\sim 4''$); it is not necessary for an object to be 
classified morphologically as a galaxy for inclusion in the ELG list. The only criterion for 
inclusion is the presence of an emission line in the objective-prism spectrum. 
In other words, there is no bias against compact, stellar appearing objects in KISS.

Although the survey data themselves provide considerable
information about the sources, obtaining follow-up spectra is an important part of the KISS project. These
spectra provide higher quality redshifts, confirm ELG candidates' status as actual ELGs, and provide the means 
by which to identify the type of activity powering the ELG. In most cases, emission lines such as 
H$\alpha$, H$\beta$, [O III] $\lambda\lambda$4959,5007, and [N II] $\lambda\lambda$6548,6583 are present
in the follow-up spectra. 
  With the higher-accuracy line strengths of 
these spectra, one can distinguish between star-forming galaxies, Seyfert 1 galaxies, Seyfert 2 galaxies, LINERs, and
QSOs. Furthermore, determination of the metallicity \citep{jason02,jason04,mmtabun,salzer04} and the star-formation rate is 
possible with accurate line-strength measurements. Follow-up spectra have been obtained for 83\% of the KR1
ELGs, 100\% of the KB1 ELGs, and 30\% of the newer KR2 ELGs at various telescopes including the Hobby-Eberly
telescope \citep{het}, the Michigan-Dartmouth-MIT 2.4m telescope \citep{mdm}, the Wisconsin-Indiana-Yale-NOAO
3.5m, the 2.1m on Kitt Peak, the Apache Point Observatory 3.5m, and the Shane 3.0m telescope at Lick Observatory \citep{jason04}.

\subsection{Surface Brightness and Half-Light Radius Measurements}
\label{kcomp}

 LCBGs typically 
are identified by their small half-light radii ($r_{hl}$) and/or their high surface brightnesses ($SB_{e}$).
 Therefore, measuring these quantities in the KISS data was an essential first step in our effort to isolate a sample of nearby 
LCBGs.
 We wrote an IRAF procedure called KCOMPACT to find $r_{hl}$ for every object in the KISS survey images. The procedure performs
a curve of growth analysis using circular apertures to determine $r_{hl}$. 
The formal uncertainty in $r_{hl}$ is estimated  by comparing the separate measurements of 
ELGs present in two survey fields (the overlap in adjacent KISS fields is approximately 3-5 arcminutes). Based on 14 objects,
 we determine the 
uncertainty in $r_{hl}$ to be 0.21 arcseconds. Combining our estimates of $r_{hl}$ with the total apparent magnitude measured during the KISS data 
processing (see $\S$ \ref{KISS_sum}) allows us to compute $SB_{e}$ using the 
following formula: \begin{equation} SB_{e} = m_{B} + 0.753 
+ 2.5\log{\pi r_{hl}^{2}} \end{equation} As the formal errors in 
the KISS apparent magnitudes are always below 0.1 magnitude, we can find the upper limit on the uncertainty of $SB_{e}$ due to
the errors in $m_{B}$ and $r_{hl}$. We find $\sigma$$_{SB_{e}}$ $<$ 0.14 mag arcsec$^{-2}$.             

\begin{figure*}[htp]
\epsfxsize=6.0in
\epsscale{0.8}
\plotone{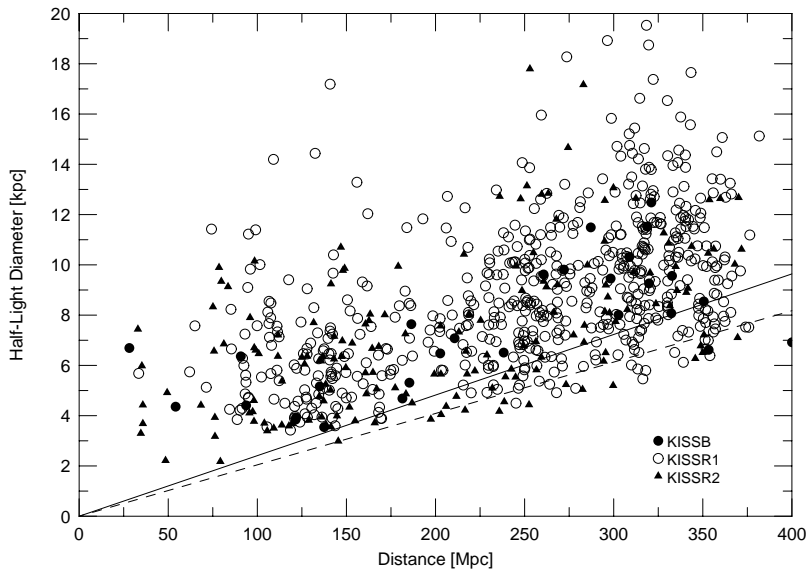}
\vskip -0.2in
\figcaption[plots/distselect2.eps]{The half-light diameter in kpc versus the distance in Mpc for a subset of
KISS star-forming galaxies from each survey strip. The dashed diagonal line represents the diameter (in kpc)
necessary for a KR2 object to be resolved at a given distance (based on the rough resolution limit discussed in 
section \ref{kcomp}). The solid diagonal line represents the
diameter (in kpc) necessary for KB1 and KR1 objects to be resolved at a given distance. Below these
diagonal lines, galaxies from their respective survey lists are unresolved. \label{fig:distsel}}
\end{figure*}

 The reliability of $r_{hl}$ and $SB_{e}$ is inextricably linked to the resolution of the image on which they are
measured. Both the image scale and the effective seeing significantly affect the accuracy of these quantities. Resolution limits 
restrict the precision with which these quantities can be determined in the more distant portion of KISS. Since \emph{all} objects 
in each survey image are analyzed
by KCOMPACT, we use the locus of the $r_{hl}$ values for objects previously identified as stars to define a rough resolution limit
for each image. Marked interactively, the resolution limit represents not the mean FWHM of the stellar distribution 
but rather is selected to encompass the entire locus of stellar objects. As we will discuss in the following 
paragraph, variable image quality across our survey images causes this limit to underestimate the true image quality for a 
modest fraction of our ELG candidates.
For KB1 and KR1, the mean resolution limit (i.e., the lower limit on $r_{hl}$) of the images is 2.50'' compared to an image scale of
2.03''/pixel; for KR2, this mean resolution limit of the images is 2.13'' compared to an image scale of 1.43''/pixel.
Therefore, objects with $r_{hl}$ under 
$2.5''$ are almost always unresolved (or, at best, marginally resolved)
 in the KISS survey direct images. Moreover, as the distance to an object increases,
so does the physical diameter that corresponds to this resolution limit. That is to say, at greater distances,
galaxies with progressively larger diameters dominate any resolved sample. Figure \ref{fig:distsel} shows this effect, plotting diameter in 
kpc vs. distance in Mpc,
 where the diagonal lines represent the nearly constant resolution limits (converted to kiloparsecs)
below which an object is unresolved. Everything but galaxies with large diameters will be 
unresolved at the redshift limit of the survey. Ultimately, this bias leads us to adopt a distance-limited sample of KISS
LCBGs (see section 2.3).

The rough resolution limit discussed in the previous paragraph suffices for a determination of an appropriate
distance-cut (see below), but fails to account for the rather extreme image quality variability of the KISS direct images. The delivered 
image quality of the Burrell-Schmidt 0.61-m at the time KISS was carried out 
could vary between $2.5''$ and $6.5''$ over a single survey field. Hence, rather than use a single value to describe the image quality 
on a given survey image, it is necessary to estimate the local value for the image quality in the vicinity of each KISS ELG. An IRAF 
script was written for this purpose. By locating the ten closest stars to each ELG and measuring the FWHM of their profiles, this script 
accounts for the variable image quality of the KISS direct images. The local FWHM (average seeing)
for each KISS ELG is the mean of the FWHM from the ten stellar profiles. As our study relies upon 
accurate measurements of $r_{hl}$ and  $SB_{e}$, we were forced to exclude those objects for which the local image quality was severely 
degraded. A ratio between the mean local
FWHM and the half-light diameter acted as our main diagnostic: if the ratio were above 1, we could not trust the measurement of
$r_{hl}$, and therefore we exclude the galaxy. In addition, all objects with the local stellar FWHM above 5'' and a diagnostic ratio 
above 0.7 were also excluded. In total 96 galaxies were cut out of the survey area, decreasing the area over which we could search for 
LCBGs by only $\sim 4\%$.  With these exclusions, we feel confident in 
the completeness of our sample. That is, in the remaining portions of KISS we should not miss any
apparently small, potential LCBG candidate simply because it is under-resolved. Of course, the galaxies remaining in our sample 
of KISS ELGs are not all well-resolved, but as we will describe, we can account for the inaccuracies in their measurements.  

\begin{figure*}[htp]
\epsfxsize=5.0in
\epsscale{0.8}
\plotone{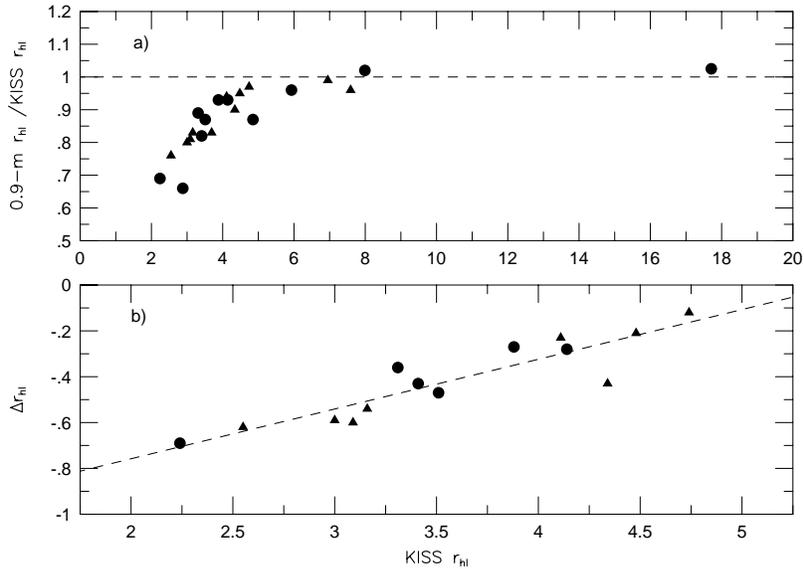}
\vskip -0.2in
\figcaption[plots/hlradcorr.eps]{ a) A plot of the ratio of the  $r_{hl}$ re-measured on the 0.9-m images
 (0.9-m $r_{hl}$) to the $r_{hl}$ measured on the KISS direct images (KISS  $r_{hl}$) versus the KISS $r_{hl}$.
The filled circles represent ELGs from KR1, and the filled triangles represent ELGs from KR2.  At an $r_{hl}$ of 
approximately 5 arcseconds, the ratio of the two quantities is very close to one. b) This plot shows the observed 
relation between $\Delta$$r_{hl}$ (0.9-m $r_{hl}$ $-$ KISS $r_{hl}$) and the KISS $r_{hl}$. Plot symbols are the 
same as in a).The linear-least-squares fit shown by the dashed line was used to apply a correction for the $r_{hl}$ 
of all KISS star-forming ELGs with  $r_{hl}$ below 5 arcseconds.  \label{fig:hlradcorr}}
\end{figure*}

In essence, our measurements of marginally resolved galaxies act as 
upper limits on $r_{hl}$ and lower limits on $SB_{e}$. 
 We explore the nature of our unresolved and marginally resolved LCBG candidates using 
higher-resolution images from the Wisconsin Indiana Yale NOAO (WIYN) $0.9$ m telescope on Kitt
Peak. In May 2003 and March 2004 we observed 
11 KISS ELGs from KR1 and 12 KISS ELGs from KR2
 under non-photometric conditions with the S2KB CCD, for which the image scale is 0.6 arcsec/pixel. 
The pre-existing KISS direct images for the galaxies, that were taken under good conditions, provided the
photometric calibration for our 0.9-m images. Despite the different image scales of KR1 and KR2, we treat them the same in the following 
analysis due to the variable image quality described above.  

For the KISS compact, star-forming ELGs in each survey strip, we re-measured the half-light
radii on the 0.9-m images 
using the same curve-of-growth routine described above,
 and compared them with the 
values of $r_{hl}$ obtained in the same way for the KISS images. The KISS $r_{hl}$
is plotted against the ratio of the 0.9-m $r_{hl}$ to the KISS $r_{hl}$ in Figure \ref{fig:hlradcorr}a to show the results of these
re-measurements. The triangles represent the ELGs from KR2, while the 
circles represent the ELGs from KR1. The relationship is equivalent for both strips:
 those objects with larger $r_{hl}$ have a ratio close to unity, while those 
objects with $r_{hl}$ near our resolution limits have ratios that deviate more considerably from unity.

We sought to quantify the correlation seen in Figure \ref{fig:hlradcorr}a
in order to develop a correction of $r_{hl}$ for all unresolved and marginally resolved ELGs in the KISS database. 
Analyzing our sets of measurements for the 0.9-m images and the KISS images,
 we can predict the way in that the difference between a galaxy's 0.9-m $r_{hl}$ and its KISS $r_{hl}$ ($\Delta$$r_{hl}$) 
increases as its $r_{hl}$ decreases below the resolution limits of KISS.
 Instead of re-observing all unresolved and marginally resolved KISS ELGs to achieve 
more accurate measurements of their $r_{hl}$, we can infer the values we would obtain. Two of the eleven KR1
 and two of the twelve KR2 ELGs observed with the 
WIYN telescope 
had $r_{hl}$ greater than 6.0 arcseconds, and were thus very well-resolved in the KISS direct images. Figure 
\ref{fig:hlradcorr}a reassures us that our measurements of $r_{hl}$ on the 0.9-m images are very close to our
measurements on the KISS images for these four
well-resolved galaxies. These four galaxies were excluded from the determination of our correction. Upon closer examination of the 
0.9-m images, one galaxy in KR1
 was observed to have a merging partner,
 making suspect any measurement of its $r_{hl}$. An additional two galaxies from each survey strip were excluded as well because they 
suffered from 
noticeably poorer image quality in the KISS direct images, skewing the relationships shown in the following plots.
 Figure \ref{fig:hlradcorr}b shows $\Delta$$r_{hl}$ versus the KISS
$r_{hl}$ for the 14 remaining galaxies. In this figure, we notice the same effect as in Figure \ref{fig:hlradcorr}a:
 apparently smaller galaxies have greater 
$\Delta$$r_{hl}$. We based the final correction for $r_{hl}$ upon the linear-least squares fit of the 14 data points
plotted in Figure \ref{fig:hlradcorr}b. The fit is shown by the dashed line. The equation for this line is: $\Delta$$r_{hl}$ $=$ 
0.22(KISS $r_{hl}$) $-$ 1.19, where $\Delta$$r_{hl}$ is in arcseconds, and the scatter about the fit is 0.11 arcseconds. 

\begin{figure*}[htp]
\epsfxsize=5.0in
\epsscale{0.8}
\plotone{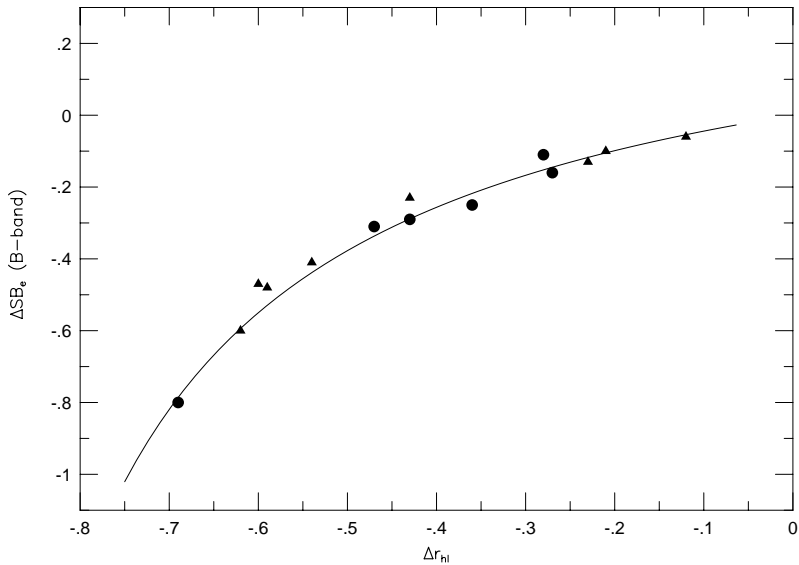}
\vskip -0.2in
\figcaption[plots/sbcorr.eps]{ The observed relation between $\Delta$$SB_{e}$ (0.9-m $SB_{e}$ $-$ KISS $SB_{e}$)
and $\Delta$$r_{hl}$ shown by the filled circles and triangles plotted with the analytical correction curve given by the equation 
$\Delta$$SB_{e}=-2.5\log\frac{r_{hl}^{2}}{(r_{hl} + \Delta r_{hl})^{2}}$. The circles represent KR1 galaxies, while the 
triangles represent KR2 galaxies.  $\Delta$$r_{hl}$ is based upon the correction for $r_{hl}$, given by the linear-least-squares 
fit shown in Figure \ref{fig:hlradcorr}. 
\label{fig:sbcorr}}
\end{figure*}

We also compared the apparent B-band magnitudes ($m_{B}$) measured on the 0.9-m and 
KISS images. Since our photometric calibration of the 0.9-m images made use of the original 
KISS calibration, our 0.9-m values for $m_{B}$ agree well with the KISS values for $m_{B}$. This re-measurement of $m_{B}$ combined
with the correction for $r_{hl}$, described above, is relevant for probing the effects of low-resolution upon the measurement of $SB_{e}$
(see equation 1).  
Figure \ref{fig:sbcorr} plots the observed relation between $\Delta$$SB_{e}$ (0.9-m $SB_{e}$ $-$ KISS $SB_{e}$) 
and $\Delta$$r_{hl}$. Since the KISS and 0.9-m
apparent magnitudes show good agreement, the difference in the measured values of $SB_{e}$ must be due to solely the change in $r_{hl}$
between the KISS and 0.9-m images. Therefore we can compute the correction for the effective surface brightness, $\Delta$$SB_{e}$,
such that it depends on only $\Delta$$r_{hl}$. \begin{equation} \Delta SB_{e} = -2.5\log\frac{r_{hl}^{2}}{(r_{hl} + \Delta r_{hl})^{2}} \end{equation}

\noindent Figure \ref{fig:sbcorr} plots the observed values of $\Delta$$SB_{e}$ and $\Delta$$r_{hl}$ for the 
14 KISS ELGs along with a curve illustrating the analytical correction given by equation 2. The observed values 
of $\Delta$r$_{hl}$ and $\Delta$$SB_{e}$ are shown by the filled triangles (KR2) and the filled circles 
(KR1).   In the equation, $\Delta$$r_{hl}$ corresponds to the linear-least-squares fit shown in Figure \ref{fig:hlradcorr}b 
for the $r_{hl}$ correction.  Figure \ref{fig:sbcorr} shows that the actual measurements of $\Delta$r$_{hl}$ and 
$\Delta$$SB_{e}$ follow the correction defined by equation 2 very well. 

We apply these corrections for $r_{hl}$ and  $SB_{e}$ to all non-excluded (see above) KISS star-forming galaxies with an
$r_{hl}$ less than 5 arcseconds. Among all KR1 star-forming ELGs with $r_{hl}$ under 5 arcseconds (i.e., those
affected by the correction), the mean change in $r_{hl}$ is -0.50 arcseconds, the mean change in diameter is -4.91 kpc, 
and the mean change in $SB_{e}$ is -0.44 mag arcsec$^{-2}$.  Among all KR2 star-forming ELGs with $r_{hl}$ under 5 arcseconds,
the mean change in $r_{hl}$ is -0.56 arcseconds, the mean change in diameter is -5.31 kpc, and the mean change in $SB_{e}$ is 
-0.57  mag arcsec$^{-2}$.

\subsection{Sample Selection}
\label{sampsel}

The ability of our local LCBG candidates to provide meaningful mass and SFR comparisons with
the higher-z LCBGs requires 
that these galaxies be genuine LCBGs. Typical methods of identifying nearby LCBG analogues usually employ cut-offs in 
surface brightness, luminosity, and color, and cannot guarantee that the galaxies are true examples of LCBGs. 
Indeed, we only can pick out those galaxies that are \emph{likely} to be LCBGs. 
Considering the nature of KISS, we believe that we are identifying 
those ELGs that possess the same properties as the LCBGs at intermediate and high redshifts, and thus, those galaxies that
are most likely to be true LCBGs. Hence, we call our galaxies LCBG candidates. 

In an attempt to match our KISS LCBG sample with the higher-redshift samples, we endeavor to employ the same
cut-offs as those used in earlier work. The sample is based on criteria proposed by 
\cite{private}, who uses cut-offs in surface 
brightness, rest-frame color, and absolute magnitude for the selection of LCBGs.  All criteria
are limited by the angular resolution of the KISS survey direct images. Certainly, LCBGs are \emph{not} AGN, and are most 
often characterized by vigorous star-formation. Therefore the sample
was selected from a subsample of KISS galaxies known from follow-up spectra to have star-formation as their primary form of activity.
All values given for $B-V$ color and
all magnitudes used in the KISS sample selection are corrected for Galactic absorption using the values of \cite{schlegel}.
Additionally, to select this sample, we used the corrected values of $SB_{e}$ determined from the analysis outlined 
in the previous section.
 
 A total of 17 KISS star-forming galaxies meet the selection criteria proposed by R. Guzm\'{a}n: 
$SB_{e}$ (\emph{B}-Band) $< 21.0$ mag arcsec$^{-2}$, $M_{B}< -18.5$, $B-V < 0.6$ with $H_0= 70$ km s$^{-1}$ Mpc$^{-1}$.
We use this absolute magnitude limit for the selection of KISS LCBGs despite our use of $H_0= 75$ km s$^{-1}$ Mpc$^{-1}$
by KISS. Since this absolute magnitude cut corresponds to $M_{B}< -18.35$ with $H_0= 75$ km s$^{-1}$ Mpc$^{-1}$, we 
apply a slightly more stringent cut for absolute magnitude by 0.15 magnitudes when we use $M_{B}< -18.5$ as one of our 
selection criteria. 
 
This sample of LCBGs is partially determined by the resolution selection 
effects described in Section \ref{kcomp} and shown in Figure \ref{fig:distsel}.
 These effects dictate that the majority of identified LCBG candidates will be located
at distances far closer than the KISS redshift limit of $z=0.095$. Considering that the median half-light diameter of the 17 LCBGs
 is 4.21  kpc, Figure \ref{fig:distsel} indicates that likelihood of finding an LCBG beyond $\sim200$ Mpc
 is low. A further implication of this effect is that any comparison of the KISS LCBG sample
with other KISS star-forming galaxies will be skewed. KISS star-forming galaxies detected at greater
distances will be, on average, considerably larger in diameter and more luminous. To account for these
effects, and to improve the completeness of our LCBG sample, we impose on it a distance limit of 185 Mpc ($z\sim0.045$).
16 of the 17 KISS LCBGs lie within this boundary, and according to the resolution thresholds marked on 
Figure \ref{fig:distsel}, any galaxy within 185 Mpc of the Milky Way 
with a diameter equal to or greater than $\sim4$ kpc will be resolved in the KISS images.
 
\begin{figure*}[htp]
\epsfxsize=5.0in
\epsscale{0.8}
\plotone{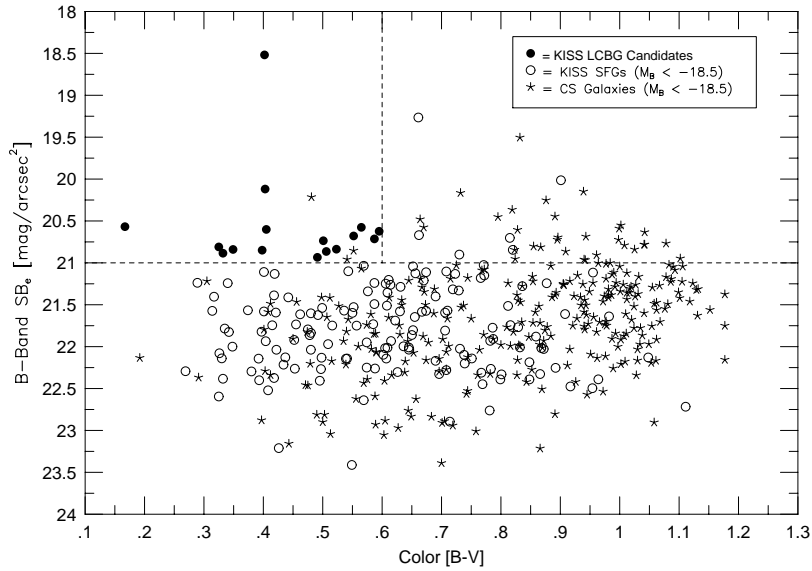}
\vskip -0.2in
\figcaption[plots/sbcolor.eps]{Effective \emph{B}-band surface brightness  plotted against B$-$V color for the KISS 
star-forming galaxies (SFGs) and Century Survey galaxies within 185 Mpc of the Milky Way. Marked by dashed lines are the 
selection cut-offs for the KISS  LCBG sample. A star indicates a galaxy from the CS, and an open circle indicates the KISS SFGs. 
We see that 16 ELGs comprise the  distance-limited LCBG sample, shown by the filled circles.\label{fig:properties}}
\end{figure*}

Figure \ref{fig:properties} shows KISS ELGs that lie within 185 Mpc of the Milky Way and have
$M_{B}  < -18.5$. The 16 KISS ELGs that remain in the sample after the application of
the distance cut-off are marked as filled circles in the plot, and the dashed lines indicate two of the selection criteria.
The stars in the plot represent a subset of ``normal'' galaxies from the Century Survey (CS) carried out by \cite{Weg01} that
lie within the distance limit and have $M_{B} < -18.5$. The Century Survey 
is a traditional magnitude-limited survey of ``normal'' field galaxies that are used here to represent the general
galaxian population. The open circles represent a comparison sample of KISS star-forming ELGs that have $M_{B}< -18.5$ and lie 
within 185 Mpc of the Milky Way.  
Henceforth, when we refer to our samples of LCBG candidates and comparison samples, we refer to the \emph{distance-limited} samples.
 It should be noted that 
all KISS ELGs selected as LCBG candidates have been identified as having star-formation as their primary form
of activity from follow-up spectra. 

\begin{figure*}[htp]
\epsfxsize=5.0in
\epsscale{0.8}
\plotone{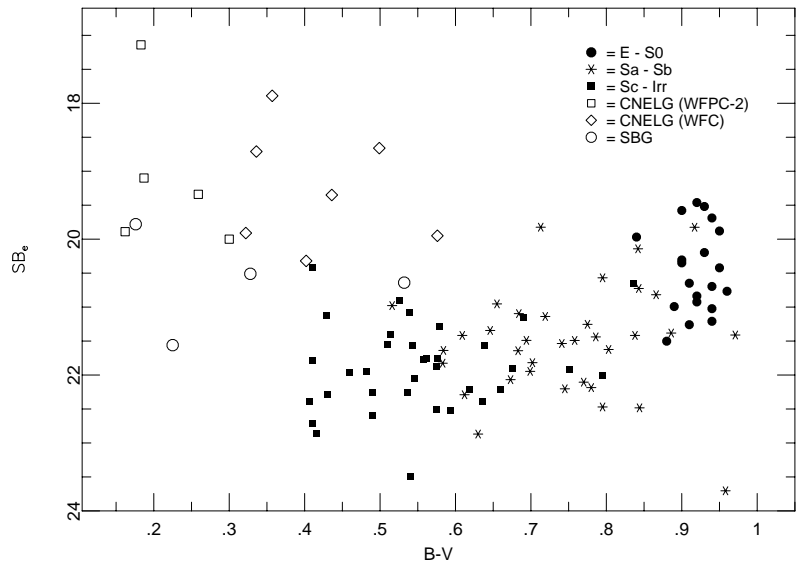}
\vskip -0.2in
\figcaption[plots/anna.eps]{Rest-frame effective \emph{B}-band surface brightness, $SB_{e}$, versus rest-frame color
B$-$V. In this plot from \cite{anna}, the CNELGs and small, blue galaxies (SBGs), also called faint, blue galaxies,
come from Koo \etal (1994,1995), while the ``normal'' Hubble-type galaxies come from a catalog assembled by \cite{frei}.
Here, the LCBGs (CNELGS and SBGs) are clearly separated from the ``normal'' galaxies. We saw a similar effect in figure 
\ref{fig:properties} for the LCBGs versus the CS galaxies. In contrast, we saw in the same figure that the 
LCBGs seem to be on the high surface brightness end of a continuous distribution of
regular star-forming galaxies. \label{fig:anna}}
\end{figure*}

 On the plot of $SB_{e}$ versus
color in Figure~\ref{fig:properties}, two features stand out. While the KISS LCBG candidates were 
selected to have surface brightnesses higher than
the typical star-forming galaxy (SFG), even the highest surface brightness LCBG candidates do not appear to be clearly
separated from the distribution of KISS SFGs. Instead, the star-forming galaxies form a more-nearly continuous distribution 
in color and $SB_{e}$, as one might expect. At intermediate redshifts, \cite{anna} find that in the 
parameter space of $SB_{e}$ and B$-$V color, LCBGs separate themselves more cleanly than in any other
parameter space from ordinary irregular, elliptical, and spiral galaxies. A plot from \cite{anna} is shown in Figure \ref{fig:anna},
illustrating this separation. 
When we compare LCBGs with the other KISS star-forming galaxies,
we do not see this effect. 
In the $z<0.045$ local universe, 
the bluest, highest surface brightness star-forming ELGs do not appear to be separate from a distribution of 
otherwise regular star-forming galaxies, but rather represent the most extreme galaxies in this class of star-forming
ELGs. However, the KISS LCBG candidates are more clearly separate from the CS
galaxies. Interesting to note is that while only three of the CS galaxies meet the selection criteria for an LCBG, 29 
CS galaxies redder than the color cut meet both the surface brightness and absolute magnitude cut. We believe these 29 
galaxies to be local, normal elliptical galaxies. Only 6 KISS ELGs fall in this region of the diagram (i.e., compact but red).

The properties of the 16 ELGs  selected as local LCBG candidates are listed in Table 1.
Columns 1 and 2 give the survey identification numbers for each galaxy (KISSR and KISSB numbers).
The apparent B-band magnitude is listed in column 3, the \emph{B-V} color in column 4, the half-light diameter in kiloparsecs
in column 5, the B-band effective surface brightness ($SB_{e}$) in column 6, the distance to the galaxy (in Mpc) in 
column 7, the absolute B-band magnitude in column 8, the logarithm of the H$\alpha$ luminosity in column 9, and the logarithm
of the 1.4 GHz radio power from \cite{jeff} in column 10. Those galaxies without a listed radio power were not detected as 
radio emitters in the Van Duyne \etal study. 

\begin{figure*}[htp]
\epsfxsize=7.0in
\epsscale{1.0}
\plotone{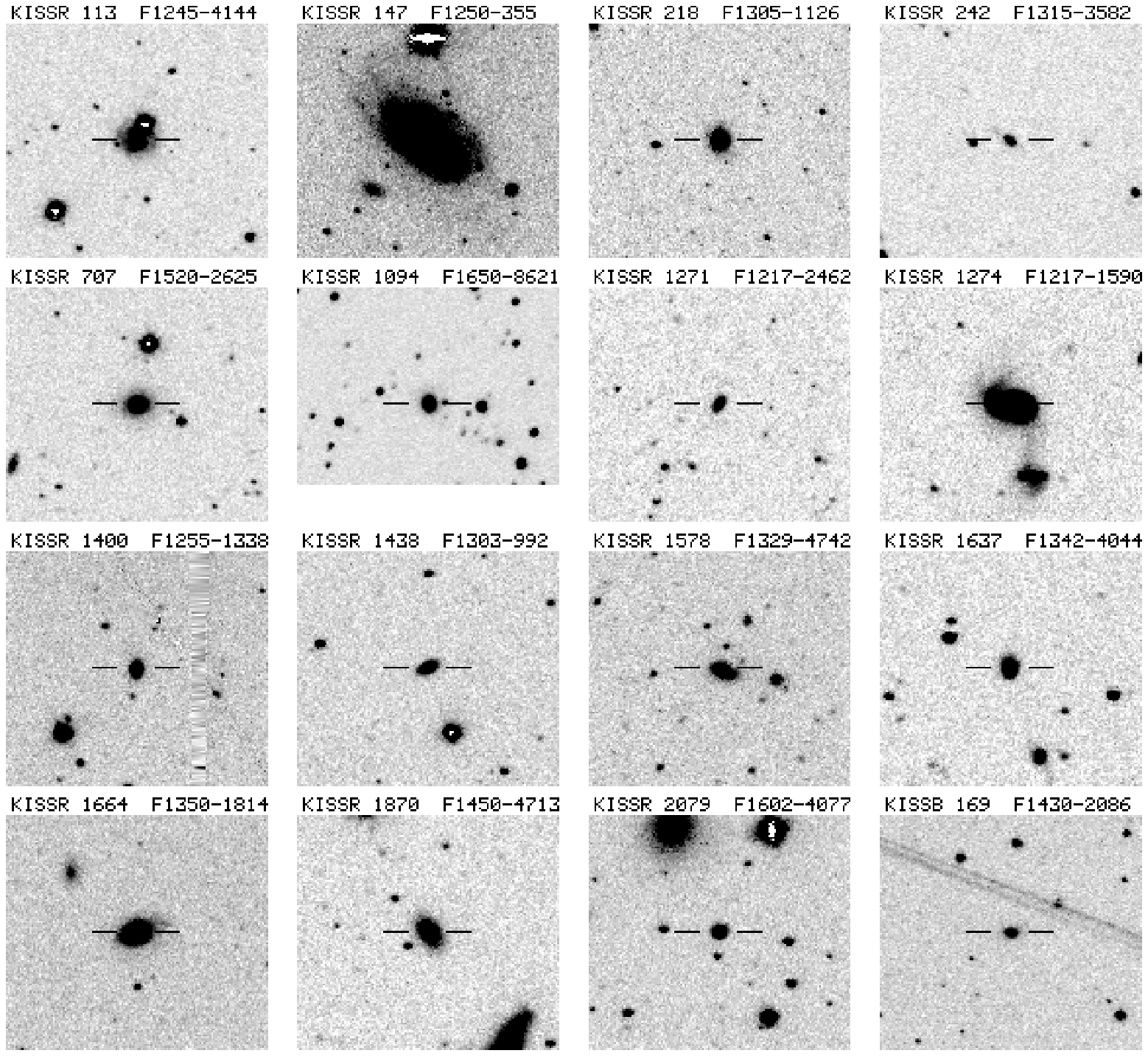}
\vskip -1.5in
\figcaption[plots/lcbgfinder.ps]{Images of all 16 LCBG candidates.  These images are derived from
the survey images, and represent a composite of the B and V filter direct images.  The LCBG candidates
are located in the center of each image, and are indicated by the tick marks. For the KR1 and KB1 
galaxies (KISSR $<$ 1128, plus KISSB 169) the images cover a FOV of 4.5 $\times$ 4.0 arcmin, while 
those from KR2 (KISSR $>$ 1128) have FOVs of 3.2 $\times$ 2.9 arcmin.  North is up and East is left in 
all images.  \label{fig:finders}}
\end{figure*}

Images of the 16 LCBG candidates made from the KISS direct images appear in Figure \ref{fig:finders}. Although the 
resolution of KISS is too low for measurements of parameters such as asymmetry and concentration, we can see
some of the morphological characteristics of our LCBG candidates on these images.  At least three of the LCBG candidates, 
KISSR 147, 1274, and 1870, appear to have disturbed morphologies, evidence for recent interactions and/or mergers. 
However, the majority of the 16 LCBG candidates appear to be small, compact, and quite symmetric, suggesting that
interactions are not the main driver of the observed star-formation activity in most cases.

\section{Properties of the Local LCBG Candidates}

\subsection{Global Characteristics}

The KISS project obtains photometrically calibrated B and V direct images along with 
the objective-prism spectra, that provide accurate measurements of each galaxy's H$\alpha$ flux.  
In addition, all of our LCBG candidates have been observed
as part of our follow-up spectroscopy program (e.g., Wegner \etal 2003; Gronwall \etal 2004a).
Hence, there is a large amount of information readily available for all KISS ELGs that allows
us to examine the properties of the KISS LCBG candidates in some detail. In this subsection, 
we present the properties of the LCBG candidates and a comparison sample of KISS star-forming 
galaxies (SFGs) with M$_{B} < -$18.5.  When the necessary data are available, we also present 
the CS comparison sample of ``normal'' galaxies that have M$_{B} < -$18.5.    All three of
these samples are taken from the same volume of space (same area of the sky, and D $<$ 185 
Mpc).  One implication of this distance limit is that the galaxies in our KISS comparison sample 
have lower luminosities and bluer colors than the overall sample of star-forming KISS galaxies. 
In general, the most distant galaxies in any sample tend to be the most luminous, and galaxies 
that are more luminous tend to be redder.

\begin{figure*}[htp]
\epsfxsize=5.0in
\epsscale{0.8}
\plotone{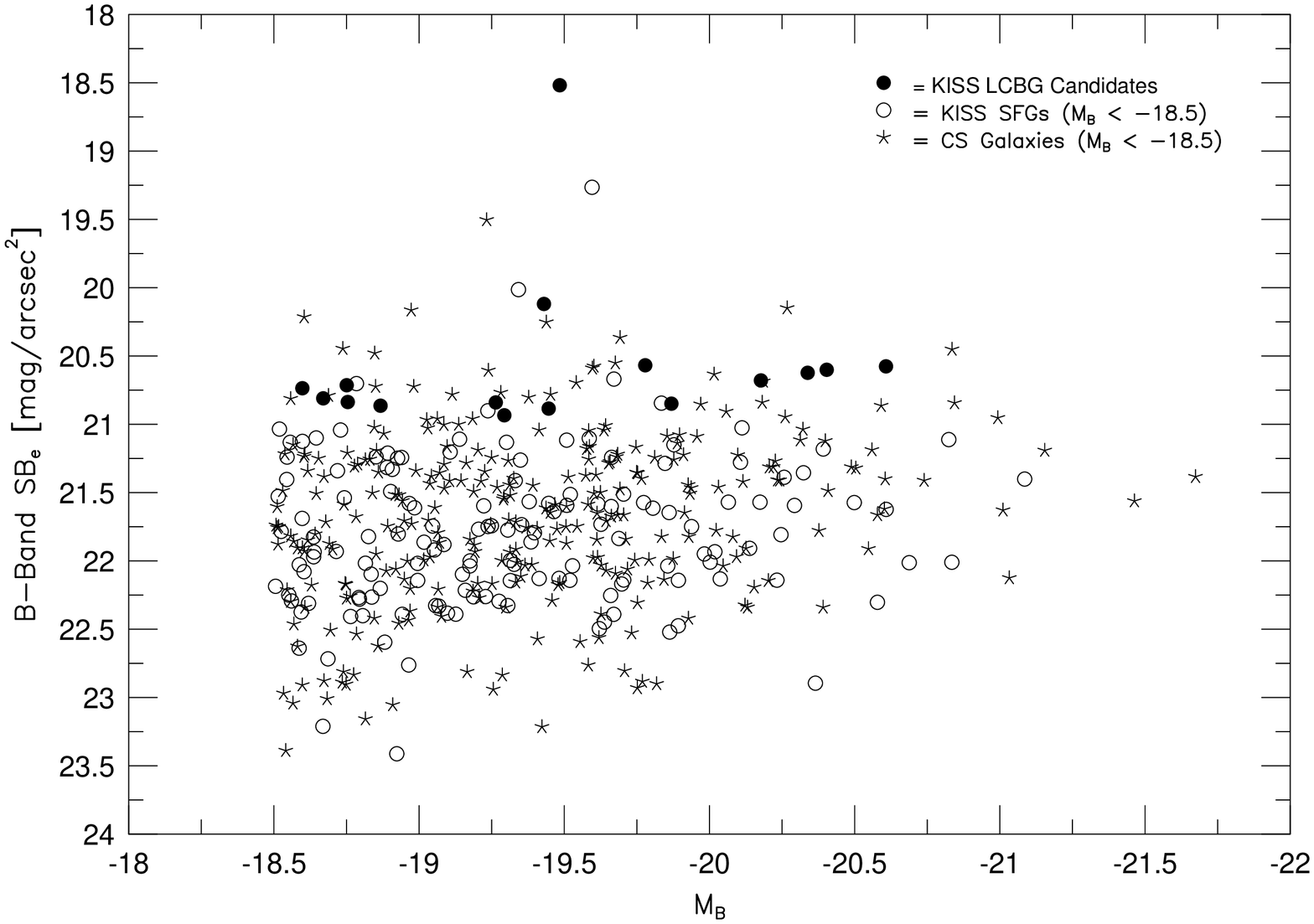}
\vskip -0.2in
\figcaption[plots/sbmag.eps]{The projection of effective \emph{B}-band surface brightness against absolute B-band magnitude 
for the KISS star-forming galaxies (SFGs) and Century Survey galaxies within 185 Mpc of the Milky Way. A star indicates a 
galaxy from the CS, and an open circle indicates the KISS SFGs.  The  distance-limited LCBG sample is shown by the filled 
circles.\label{fig:sbmag}}
\end{figure*}

The B-band absolute magnitudes of the KISS LCBGs, KISS star-forming galaxies, and CS 
magnitude-limited sample are plotted against SB$_{e}$ in Figure \ref{fig:sbmag}.  Together
with Figure \ref{fig:properties}, these two plots exhibit the selection criteria of the LCBG candidates,
and also illustrate nicely the properties of the three galaxy samples.  As pointed out in the previous
section, the KISS SFG comparison sample exhibits a fairly continuous distribution in all three
parameters -- SB$_{e}$, M$_B$, and B$-$V color -- with the LCBG candidates representing the
extreme cases in terms of color and surface brightness (Figure~\ref{fig:properties}).  The CS and
KISS SFG samples overlap substantially in both Figures~\ref{fig:properties} and \ref{fig:sbmag}, although the
latter are on average much bluer.   The LCBG candidates are seen to be distributed fairly evenly
in absolute magnitude between $-$18.6 to $-$20.6.  They define the upper envelope of the KISS 
ELGs in Figure~\ref{fig:sbmag}. 

\begin{figure*}[htp]
\epsfxsize=5.0in
\epsscale{0.8}
\plotone{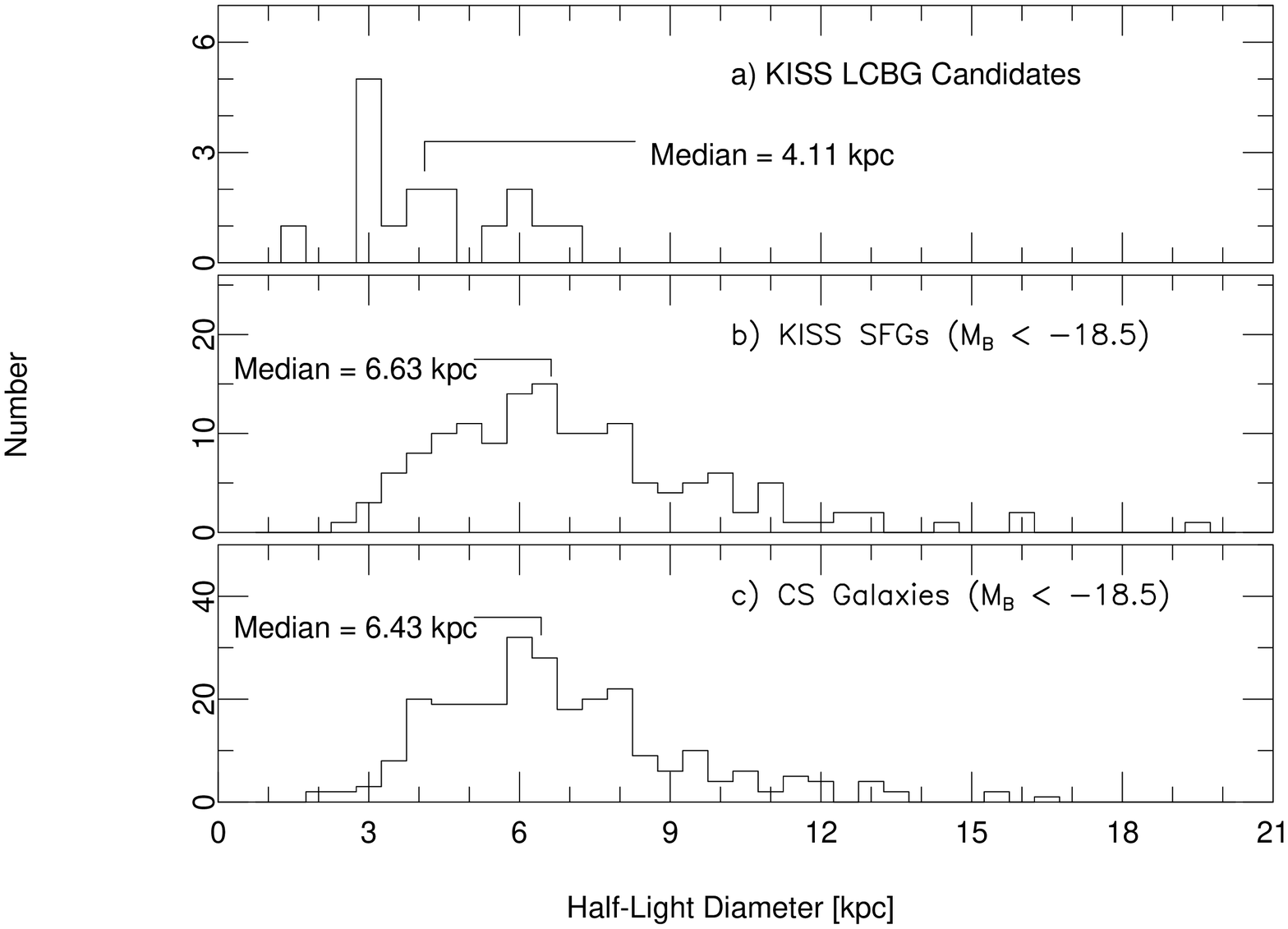}
\vskip -0.2in
\figcaption[plots/dlhistdiam.eps]{Distribution of the derived diameters in kiloparsecs for the 16 KISS
LCBG candidates, the comparison sample of KISS SFGs, and the comparison sample of CS galaxies. 
The median diameter of the LCBG sample is 4.11 kpc, the median diameter of the KISS comparison sample is 
6.63 kpc, and the median diameter of the CS comparison sample is 6.43 kpc.  \label{fig:histdiam}}
\end{figure*}

Figure \ref{fig:histdiam} shows histograms of the half-light diameters of the LCBG candidates, the 
KISS comparison sample SFGs, and the CS comparison sample ``normal'' galaxies.  The median
diameters for the three samples, 4.11 kpc, 6.63 kpc, and 6.43 kpc, respectively, are indicated in
the figure.   As expected, the KISS LCBGs are, on average, 60\% smaller than the CS galaxies and 
SFGs.  The largest LCBGs have sizes comparable to the medians for the other two distributions.  
Since the samples have comparable luminosities, the difference in diameter accounts for most of the 
difference in SB$_{e}$. 

\begin{figure*}[htp]
\epsfxsize=5.0in
\epsscale{0.8}
\plotone{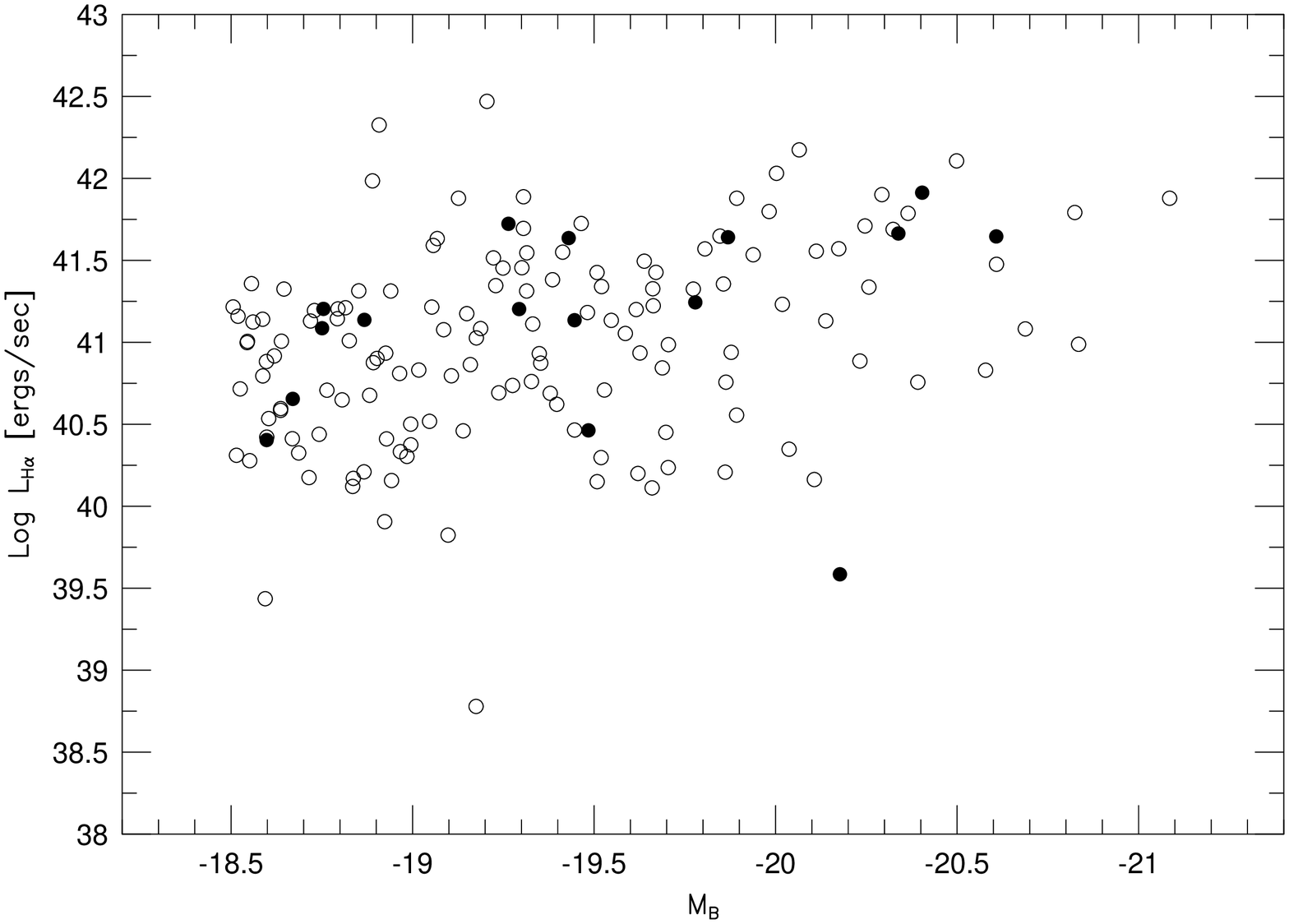}
\vskip -0.2in
\figcaption[plots/halum.eps]{ H$\alpha$ luminosities are plotted versus absolute B-band magnitudes for the 
16 KISS LCBG candidates (filled circles) and the comparison sample of KISS SFGs (open circles).  The median 
Log $L_{H\alpha}$ of this LCBG sample is 41.21 ergs/sec, which is approximately 1.6 times more luminous than 
the median Log $L_{H\alpha}$ of the comparison sample, 41.01 ergs/sec. \label{fig:hamag}}
\end{figure*}

Figure \ref{fig:hamag} shows the H$\alpha$ luminosity plotted against  M$_B$ for the two KISS 
samples.  The sample of 16 KISS LCBG candidates exhibits a higher median H$\alpha$ luminosity by 
a factor of 1.6 than the comparison sample SFGs.  Because we consider only actively star-forming 
galaxies in both samples, we can draw an important conclusion from this result.  In general, the 
H$\alpha$ luminosity indicates the activity level of a galaxy, telling us how many ionizing stars a 
galaxy contains.  The KISS LCBG candidates have, on average, 60\% higher current star-formation 
rates than the less compact SFGs.  There is a clear tendency for the LCBG candidates to be located
near the top of the diagram, again illustrating that they are above average relative to the comparison
sample.  Note that not all of the LCBG candidates are in the high star-formation rate category.  One 
of the KISS ELGs with the {\it lowest} H$\alpha$ luminosity, KISSR 147, happens to be included 
among the LCBGs.  We note in passing that the measurement of the H$\alpha$ flux in this object may
well be underestimated, since it is so extended on the sky and since the star-formation appears to
be spread over a large fraction of the galaxy.  The KISS objective-prism flux measurement is made over 
an 8 arcsec wide region centered on the galaxy.  Hence, our current estimate for the H$\alpha$ luminosity 
of KISSR 147 may be significantly lower than the actual value.  We do not believe that this
problem occurs for any of the other KISS LCBGs.

\begin{figure*}[htp]
\epsfxsize=5.0in
\epsscale{1.0}
\plotone{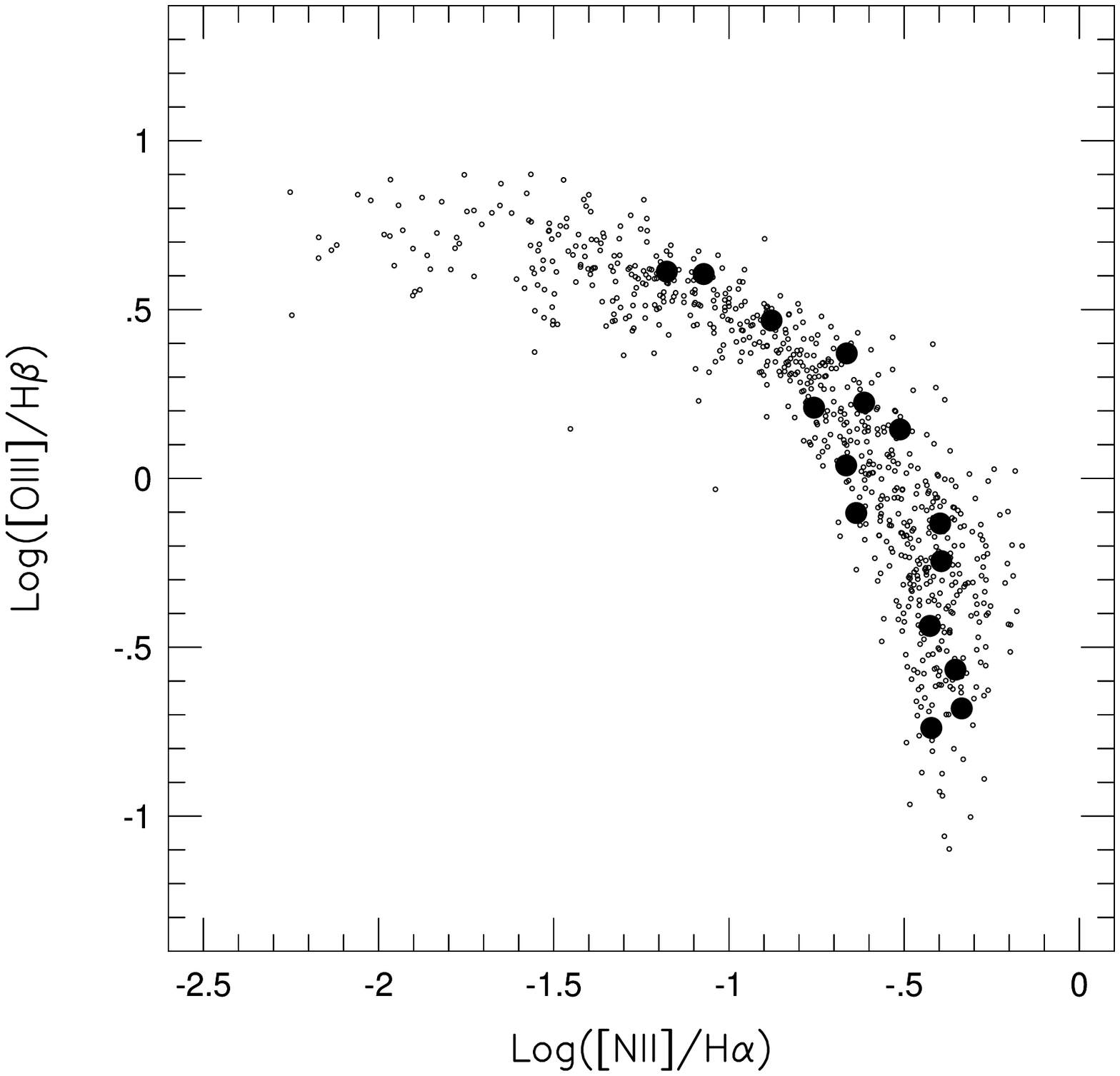}
\vskip -0.2in
\figcaption[plots/diagplot.eps]{ A plot of the logarithm of the [\ion{O}{3}]/H$\beta$ ratio vs. the logarithm of 
the [\ion{N}{2}]/H$\alpha$ ratio for all KISS SFGs. The large filled 
circles are the LCBG candidates. The characteristic \ion{H}{2} sequence is shown by the KISS SFGs where the low 
luminosity, metal poor galaxies fall in the upper left of the diagram, and the high luminosity, metal rich galaxies fall in 
the lower right region. The KISS LCBGs do not appear to deviate from this sequence.\label{fig:diagplot}}
\end{figure*}

It is also relevant to consider the spectral properties of the LCBG candidates.  Do their spectra
reveal any systematic differences relative to the less compact SFGs of the KISS sample?  We try
to explore this question in two ways.  First, in Figure~\ref{fig:diagplot}, we plot an emission-line
ratio diagnostic diagram showing the locations of all of the KISS SFGs (regardless of distance).
We plot the logarithms of the [\ion{O}{3}]/H$\beta$ ratio vs. the [\ion{N}{2}]/H$\alpha$ ratio.
The smaller open circles are the ``regular" star-forming galaxies, while the larger filled symbols
indicate the location of the 15 LCBG candidates with measured line ratios.  
The diagram reveals the characteristic \ion{H}{2}
sequence, where the SFGs fall along an arc in the diagram stretching from the upper left
(typically lower luminosity, lower metallicity galaxies) to the lower right (typically more luminous,
higher metallicity objects).  The LCBGs are distributed fairly uniformly along the lower 2/3rds
of the arc.  Our selection against lower luminosity galaxies prevents them from falling in the
upper left portion of the \ion{H}{2} sequence.  We detect no obvious trends amongst the LCBGs
that would lead us to believe that their spectral properties are systematically different from the
rest of the KISS SFGs.

\begin{figure*}[htp]
\epsfxsize=5.0in
\epsscale{0.8}
\plotone{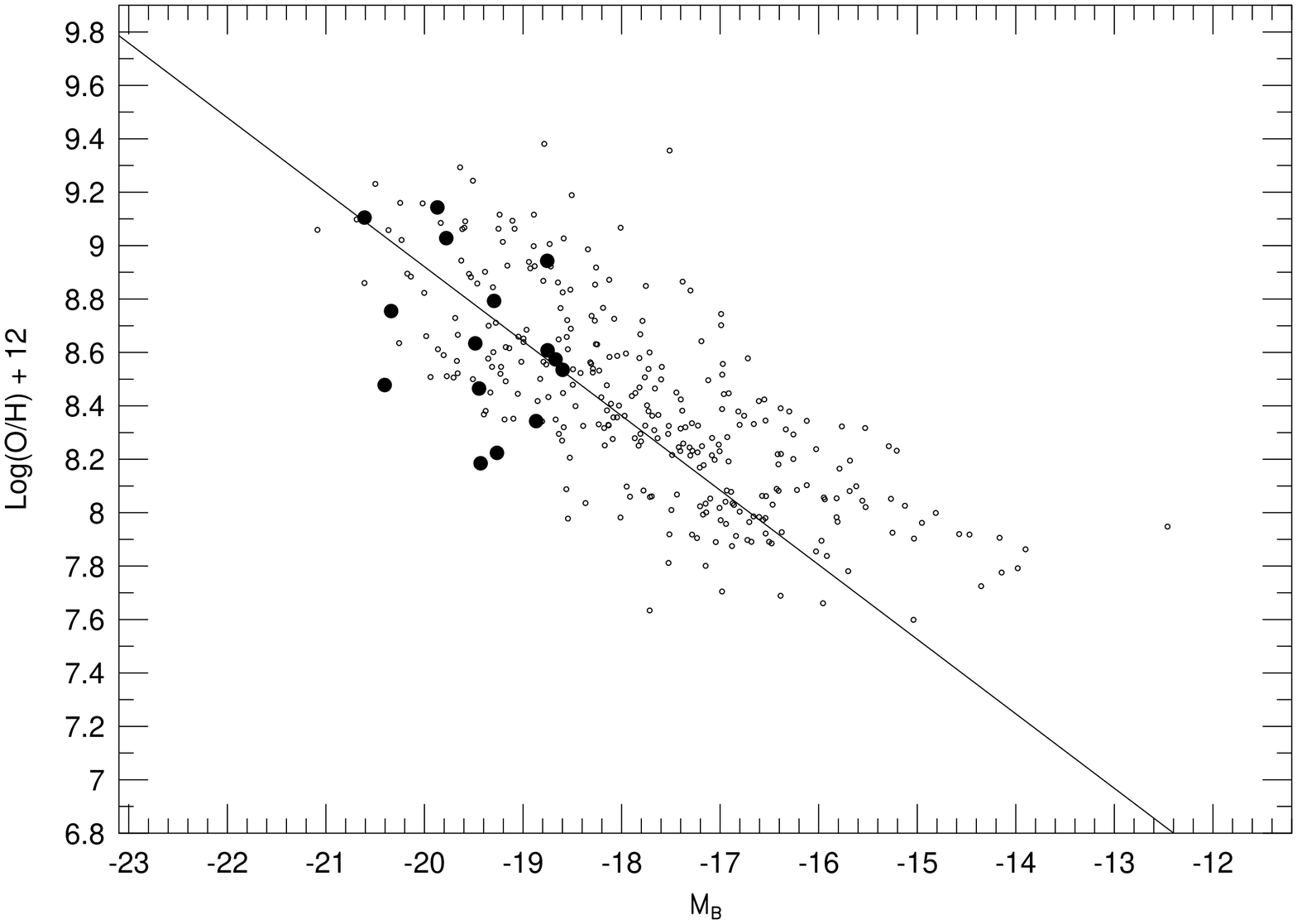}
\vskip -0.2in
\figcaption[plots/abun.eps]{$M_{B}$ versus the metal abundance (Log (O/H) + 12) for KISS SFGs within 185 Mpc of the 
Milky Way. The 15 LCBG candidates with derived abundances are the filled circles on this plot, and the KISS SFGs are 
the smaller circles. The line represents the linear L-Z fit for KISS galaxies derived in \cite{jason02}.\label{fig:abun}}
\end{figure*}

We also consider the metal abundances (log(O/H) + 12) of these 15 LCBG candidates compared to the 
sample of all KISS SFGs within 185 Mpc of the Milky Way.  Using the linear luminosity-metallicity (L-Z) 
relation of \cite{jason02}, we plot metal abundance versus absolute magnitude for the 
KISS SFGs.  Figure \ref{fig:abun} shows a weak tendency for the LCBGs, shown 
as filled circles on the plot, to lie at or below the mean L-Z relation.  The mean abundance of the 15 LCBGs 
is 8.65, compared to the mean abundance of the comparison sample (KISS SFGs with D $<$ 185 Mpc
and M$_{B} < -$18.5) of 8.75.  The statistical significance of this 0.1 dex difference is minimal, since the formal error
in the difference of the two means is 0.08.  Still, this small difference may suggest that either the LCBGs are under-producing
heavy elements compared to the comparison SFGs, are systematically losing a larger fraction of their metals 
via outflows, or the enhanced star-formation of the LCBGs is producing a modest luminosity enhancement 
relative to the rest of the SFGs.  While it is not possible to definitively select between these options, the
high current star-formation rates exhibited by the LCBGs would appear to be inconsistent with the first
of these choices, while the higher central mass densities implied by the high SB$_e$ of the LCBGs
is inconsistent with the second (see discussion in \cite{salzer04}).  Hence, we conclude that the LCBG 
candidates appear to be offset slightly to the left of the KISS SFGs, indicating that their luminosities may not be
representative of their masses.  While this latter point is likely true in  general, for all star-forming galaxies 
(e.g., Lee \etal 2004), it appears to be even more the case for the LCBG sample.

\subsection{Radio Continuum Emission}
\label{radio}

To some extent, all galaxies exhibit radio emission, but it is those that are the most radio-luminous
that are known historically as ``radio-galaxies.'' Because star-forming galaxies are often radio sources, 
we investigate the radio properties of the LCBG candidates as compared to a sample of all KISS
SFGs, expecting to detect radio emission for some subset of our samples.  All radio fluxes cited in this 
work come from \cite{jeff}, who used the FIRST \citep{first} and NVSS \citep{nvss} 20-cm radio continuum
surveys to study the radio emission associated with the 2158 KISS ELGs from KR1 and KR2.  Because that 
study did not include the KB1 portion of KISS, we exclude the KB1 ELGs from this part of our analysis. 

13.7\% of all KISS ELGs known to be star-forming galaxies in the KR1 and KR2 survey strips are detected
by FIRST and NVSS in the radio.  This is an upper limit to the fraction of radio-detections, however, since
only 57.6\% of the KR1 and KR2 galaxies currently have follow-up spectra (note that {\it all} of the radio-detected 
KISS galaxies possess follow-up spectra).  If we assume that $\sim$90\% of the galaxies lacking follow-up
spectra are also SFGs, then the fraction of SFGs with radio continuum detections drops to only 7.4\%.
This latter value is likely to be closer to the actual percentage of radio-detected SFGs.  We compare this 
number to the fraction of radio galaxies within our LCBG candidate sample (from only KR1 and KR2) and to 
the fraction of radio galaxies within our distance-limited comparison sample of KR1 and KR2 SFGs with 
M$_{B}< -$18.5.  Fully 53\% of the 15 LCBG candidates have detectable radio continuum emission, while 
28\% of the comparison sample galaxies have detected radio fluxes. 

The LCBGs clearly have a stronger propensity to possess detectable radio emission.  We hypothesize that 
the intense star-forming activity indicated by the high  $SB_{e}$ of the LCBG candidates is the primary 
contributor to this effect.  In star-forming galaxies, the energy given to the relativistic electrons which 
cause synchrotron radiation comes from the supernova explosions occurring over the past $\sim$3$\times$10$^{7}$ 
years \citep{cram}.  Presumably, the KISS LCBGs produce more supernova in this time period then regular 
star-forming galaxies due to their above-average star formation rates (previous section).  Additionally, due 
to their compact nature, LCBGs may have stronger magnetic fields than regular star-forming galaxies.  These 
properties, in concert, enhance the detectability of synchrotron radiation in LCBGs relative to other star-forming
galaxies.

Despite the high proportion of radio detections among the LCBGs, their radio powers do not stand
out.   For the eight detected LCBGs, the mean radio power is 10$^{21.81}$ W Hz$^{-1}$.  This is
statistically indistinguishable from the mean of the distance-limited comparison sample (10$^{21.84}$ 
W Hz$^{-1}$), and significantly less that the average of all KISS star-forming galaxies detected
by FIRST and NVSS (10$^{22.11}$ W Hz$^{-1}$).  The majority of the overall radio-detected SFGs
lie at distances beyond 185 Mpc, so it is no surprise that their radio luminosities are higher.  However,
it is perhaps a bit surprising that the LCBGs do not have higher radio powers than the comparison
SFGs.  Of course, the number of LCBGs in this analysis is small, so it may be premature to draw strong
inferences regarding the radio powers at this time.  What is clear, however, is that the LCBGs are 
more commonly detected as radio emitters than are the overall population of star-forming galaxies
in the same volume.

\subsection{Volume Density of Local LCBGs}
\label{vd}

The space density of the intermediate and high redshift LCBGs compared with that of low redshift 
LCBGs acts as an essential constraint on their density evolution.  Nearly all previous studies of LCBGs 
stress the apparent extreme density evolution of LCBGs.  For our sample of LCBG candidates with 
z $<$ 0.045 we compute the volume density using the 1/$V_{max}$ method.  A complete description of 
the method along with the specific completeness parameters used in the calculations can be found in 
\cite{kisscomplete}.  In brief, we first determine whether the detectable volume of an individual galaxy is 
limited by the effective volume set by the KISS redshift limit or by its own emission-line flux. Then, we 
sum the inverse of the maximum volume within which each galaxy is detectable for all galaxies in this sample. 
Accordingly, we determine the space density of local LCBGs to be $5.4\times10^{-4}h_{75}^{3}Mpc^{-3}$. 
In order to arrive at an estimate of the uncertainty in the volume density, we will assume that the main source 
of error is the Poissonian uncertainty in the number
of LCBGs (i.e., $\sigma$ = $\sqrt{N}$).  This is reasonable, since the uncertainties
in the line flux completeness limits and the relevant volumes are fairly small.
Hence, since there are 16 LCBGs in the distance-limited sample, the computed
volume density carries with it a 25\% uncertainty.

A handful of previous studies have derived estimates for the space densities of
LCBGs at intermediate and high redshifts. For example, \cite{phil97} find the volume density of 
LCBGs at redshifts between 0.4 and 0.7 with M$_{B} < $ -18.5 to be $2.2\times10^{-3}h_{75}^{3}Mpc^{-3}$, 
and that of LCBGs at reshifts between 0.7 and 1.0 with  M$_{B} < $ -18.5 to be $8.8\times10^{-3}h_{75}^{3}Mpc^{-3}$.
\cite{low97} estimate a total space density of the high redshift ($2.2<z<3.5$) population of LCBGs to be
$3.7\times10^{-3}h_{75}^{3}Mpc^{-3}$. This latter density estimate must be taken as a lower limit, 
since the Lowenthal \etal sample is at such high redshifts that only the most luminous sources are
detectable.  A correction to account for the incompleteness at the lower luminosities
would likely raise the density estimate substantially.  A large (and probably rather
uncertain) incompleteness correction to account for missing lower luminosity LCBGs was applied by 
Phillips \etal to their higher redshift subsample.  Because of the large uncertainties associated with the LCBG
density estimates in these higher redshift samples, one should use caution in interpreting the numbers.  We 
estimate that the uncertainties in these densities, including any errors associated with the incompleteness 
corrections, must be at least 33\% for the lower redshift value of Phillips et al., and closer to 50\% for
the two higher redshift estimates.  

The uncertainty that arises from different absolute magnitude cut-offs at different redshifts can be alleviated 
somewhat by applying a {\bf brighter} cut-off to our sample.  By making our local sample's luminosity limit consistent
with the absolute magnitude limit of the higher-z samples, rather than extrapolating the high-z absolute magnitudes 
to lower luminosities  (as was done by \cite{phil97} as described above), one avoids introducing some large and
very uncertain corrections.  For example, \cite{phil97} had an absolute magnitude limit of -20 (using H$_{o} = 50$ 
km s$^{-1}$ Mpc$^{-1}$) for their sample at $0.7<z<1.0$, which corresponds to an absolute magnitude limit of 
$-19.12$ for H$_{o} = 75$ km s$^{-1}$ Mpc$^{-1}$.  We derive a volume 
density of $3.9\times10^{-4}h_{75}^{3}Mpc^{-3}$ for our sample of local LCBGs with when we adopt M$_{B} = -$19.12
as our luminosity limit.  For comparison, \cite{phil97} find the volume density of LCBGs at reshifts between 0.7 
and 1.0 with  M$_{B} < $ -19.12 to be $4.05\times10^{-3}h_{75}^{3}Mpc^{-3}$.  We note that this method is impractical 
to use for the Lowenthal \etal sample, and unnecessary when comparing to the lower-redshift Phillips \etal sample.

With the above caveats in mind, we compare the volume densities of LCBGs from our
local (z = 0) sample to those of the higher redshift studies. The value \cite{phil97} obtain for their lower 
redshift sample implies that the volume density of LCBGs has declined by a factor of $\sim$4 between z $\sim$ 0.5 
and the present, while their higher redshift  value suggests that the density of LCBGs has dropped by a factor of 
$\sim$16 between z $\sim$ 0.85 and today.  This latter factor becomes $\sim$10 for the comparison using the volume 
density of the KISS LCBGs  with M$_{B} < $ -19.12.
The density drop implied by \cite{low97} for the high reshift LCBGs ($2.2<z<3.5$) is at least a factor of $\sim$7, and likely
to be much higher. The work of \cite{lilly96,phil97,lilly98} and \cite{cfrs} all suggest that the volume density
of LCBGs drops off by a factor of $\sim$10 from $z=1$ to the present. This prediction is in general agreement with 
the results presented here, given the large errors associated with the density estimates. That is, the space density 
derived for our sample of LCBGs compared with that of the higher redshift samples supports the claim that these galaxies 
are indeed a strongly evolving population. However, the uncertainties associated with the volume density of LCBGs at 
higher redshifts precludes an accurate assessment 
of exactly how extreme their evolution is. Hopefully, future studies of high redshift LCBG populations will
constrain their volume densities more definitively, and lead to a better understanding of their evolution
with look-back time.

\section{Discussion}

This paper deals with the problem of LCBGs by approaching only one of its facets. Namely, we have sought to identify
local examples of LCBGs in an effort to constrain
the properties of those LCBGs present in such profusion at intermediate- and high-$z$. However, we must be cautious in
accepting that our sample is composed of genuine LCBGs $-$ indeed, we call the galaxies in our sample LCBG candidates.
The classification scheme for LCBGs presented by Jangren \etal defines a 
six-dimensional parameter space in which LCBGs are effectively isolated from other types of galaxies. However, we could
not apply this method for our LCBG selection, as KISS direct images  
did not have the resolution required to perform the prescribed 
tests. And, while Jangren \etal conclude that the parameter space of B$-$V color vs. $SB_{e}$ is the best
at separating the LCBGs from other types of galaxies,
 they acknowledge that there is still no single cut that will eliminate all non-LCBGs nor 
include all \emph{bona fide} LCBGs. 

Specifically, employing a color limit to select local LCBGs based on the corrected rest-frame colors of higher-redshift 
LCBGs is a bit problematic. The many corrections that must be applied to color, (i.e., K-corrections, Galactic reddening, internal extinction, 
emission-line contamination) make it a difficult parameter to use for defining a sample of galaxies. Additionally, the 
rest-frame colors of star-forming galaxies at high redshift will be inherently biased to be
bluer than those of lower redshift star-forming galaxies.  This is because host galaxies at higher redshift have not had as much time 
to assemble as those in the nearby universe.  Hence, they will be intrinsically less luminous, and a star-formation 
event at high-$z$ will tend to dominate the galaxian color more than at lower redshifts. Accounting for this effect, however, 
would be a difficult task, mainly because of the uncertainty in our understanding of galaxy evolution. 

The necessity of a color parameter, however, is manifest in Figure \ref{fig:properties}.  Luminous elliptical galaxies from the 
Century Survey meet both the luminosity and surface brightness criteria for LCBGs.  However, LCBGs are not similar to 
local elliptical galaxies. A color cut acts as the means for excluding early-type galaxies in traditional, 
magnitude-limited surveys.  By contrast, in emission-line galaxy surveys such as KISS that contain few, if any, elliptical galaxies, the 
color criterion becomes less pertinent. Limiting the samples of LCBGs to actively star-forming galaxies should ensure their 
relative blueness.  While not extremely blue 
compared to other star-forming galaxies, the KISS LCBG candidates are considerably bluer than ``normal'' galaxies. The
Century Survey ``normal'' field galaxies within the distance limit of 185 Mpc 
have a median B$-$V color of 0.77, considerably higher than the median color of the LCBG sample, 0.44. 
Were we to have selected a sample of LCBGs using only the luminosity and surface-brightness criteria, and limiting our sample 
to only actively star-forming galaxies, our sample
would have contained 22 KISS LCBG candidates with a median  B$-$V color of 0.53, still considerably bluer than the CS galaxies. 

We select our local samples of LCBGs from only actively star-forming galaxies not only as a way to ensure their relative blueness, 
but also to further constrain their 
properties to match those of the higher-redshift samples. Indeed, all known LCBGs at high-$z$ are vigorously forming stars,
as evidenced by their strong emission lines. This property gives rise to important questions regarding the nature of LCBGs. Is 
active star-formation always present in galaxies meeting the LCBG selection 
criteria? Can an LCBG be an LCBG without its starburst? What do LCBGs look like when and if the star-formation fades? Do the 
LCBG criteria allow for non-active galaxies? 

In an attempt to address these questions, we looked at the galaxies located in the KISS database that are $not$ ELGS, but rather
galaxies detected by the Century Survey that have measured redshifts, colors, and absolute magnitudes. Since the
first two survey lists of KISS (KR1 and KB1) overlap parts of the CS, we possess direct images taken through B and V 
filters of essentially all of the CS galaxies.  By combining our imaging photometry with the CS redshifts
\citep{Weg01}, we are able to carry out the same analysis on the CS galaxies as we previously did for the KISS ELGs.
Applying the same cut-offs in $SB_{e}$, $M_{B}$, and B$-$V color to these galaxies yields a sample of seven CS LCBG candidates.
Of these, only three were not also detected by KISS.  All three lie outside the area of sky covered by 
KR1. One of these galaxies has been identified using the NASA Extragalactic Database\footnote{This research has made 
use of the NASA/IPAC Extragalactic Database (NED) which is operated by the Jet Propulsion Laboratory, California Institute 
of Technology, under contract with the National Aeronautics and Space Administration.} as a nearby Seyfert 2 with very weak 
emission lines. KISS could not have detected this galaxy due to its strong continuum and the weakness of its H$\alpha$ line. 
The others were initially detected by the Case Low-Dispersion Northern Sky Survey \citep{cw}, with published follow-up spectroscopy 
in \cite{john95}. One galaxy, CG 60, is a star-forming galaxy  with  strong H$\alpha$ and weak \ion{O}{3}. The other, 
CG 90, appears to be a weak-lined star-forming galaxy, possibly a post-starburst system. This latter galaxy appears to be
the closest thing to a ``normal'' (i.e., not actively star-forming) galaxy in the $\sim$100 square degrees of the Century 
Survey that meets the LCBG selection criteria.  However, even CG 90 has {\it some} current star-formation.  The study by 
\cite{drinkwater} which used the Fornax Spectroscopic Survey to find a sample of bright, compact 
galaxies that are unresolved on photographic plates obtains a similar result. Of their 13 local ($z<0.21$) galaxies that
resemble the higher-$z$ LCBGs, they find only four that do not have strong emission lines. One of the galaxies has a 
post-starburst spectrum and another has a spectrum indicating an old population.

This paucity of non-active LCBGs suggests that they are, in general, extremely rare, and that perhaps the star-formation that 
seems always to be 
present in LCBGs is a key factor to getting labeled as an LCBG in the first place.  Clearly the star-formation activity helps
to make the LCBG.  The starburst will both increase the central luminosity of the galaxy, and lower the half-light radius r$_{hl}$.
These two effects combine to increase SB$_e$ and lower the color.  Is a central starburst {\it required} for a galaxy to become
an LCBG?  Perhaps not, but it certainly appears to be strongly favored.

While it is possibly premature to attempt to resolve the questions regarding the evolution of LCBGs discussed in \S 1, we can at least 
use the properties of the local candidates to speculate a bit.  The distribution of the KISS ELGs in Figure~\ref{fig:properties} suggests
that they exhibit a continuum of physical characteristics in the surface brightness, color, and luminosity parameter space.  Galaxies
classified as LCBGs using the Jangren \etal selection criteria represent those galaxies that currently lie at the extreme end of that
continuum.  Clearly, however, there are plenty of KISS ELGs that are located just below the surface brightness threshold
that distinguishes our LCBG candidates from the remainder of the star-forming galaxy population.  Since the KISS LCBGs do,
on average, exhibit higher H$\alpha$ luminosities, it is likely that at least some of them will evolve out of the LCBG
region in Figure~\ref{fig:properties} as their starbursts fade.  Will they fade substantially, as proposed by \cite{koo95} and
\cite{guz98}, dropping in luminosity by 4 magnitudes or more, becoming substantially redder and more gas poor and eventually
resembling dwarf elliptical galaxies?  

We feel that it is unlikely that any of our LCBG candidates will evolve into early-type galaxies.
The evolutionary scenario proposed by \cite{koo95} and \cite{guz98} requires that the gas present in the LCBG
be removed, presumably by strong supernovae-driven outflows.  Hydrodynamic models (e.g., Mac Low \& Ferrara 1999) suggest that
this complete blow-away of a galaxy's ISM does not occur for any but the least massive systems.  Therefore, we do not expect
any of our current sample of LCBGs to become gas-poor and fade into red ellipticals.  Of course, it remains possible
that the higher redshift LCBGs represent a more heterogeneous sample of objects \citep{guz97,phil97,hammer,pisano,garland}.
If some of the high-$z$ LCBGs did lose their gas and evolve into dwarf elliptical galaxies by the present time, our selection
method would not be sensitive to them.  However, based on the models of Mac Low \& Ferrara, we would expect that the
majority of LCGBs should retain their gas to the current epoch.  This appears to be the case with our sample.

The continuum of physical characteristics evident in Figure~\ref{fig:properties} suggests that luminous galaxies possess
a range in their central surface brightnesses / central mass densities.  This has previously been observed in dwarf galaxies
(e.g., Papaderos \etal 1996, Salzer \& Norton 1999), where the most actively star-forming dwarf galaxies (a.k.a. blue compact
dwarfs, or BCDs) are seen to lie at the extreme end of a continuum that stretches to very low central surface brightnesses 
at the other extreme.  Furthermore, the gas distributions in the BCDs are seen to be much more centrally concentrated than 
those in more quiescent dwarf irregulars (e.g., van Zee, Skillman \& Salzer 1998).
The combination of high central mass densities and high central gas densities make BCDs particularly efficient at forming stars.
We assume that this is no accident: BCDs are by their nature well equipped to be making stars.  By analogy, one might argue that
LCBGs are predisposed to make stars efficiently because they too represent the extreme high surface brightness, high central mass
density end of a continuous distribution, albeit for more luminous galaxies.  In this picture, LCBGs are preferentially in a
star-forming phase for a large fraction of the time.  When not undergoing a strong starburst, they will fade somewhat (mostly in SB$_e$, but
also slightly in color and luminosity) and join the population of galaxies lying below the LCBG selection region in 
Figure~\ref{fig:properties}.  This would help to explain the lack of {\it any} non-star-forming LCBGs present among the CS
galaxies.  Even when not strongly bursting, post LCBGs may exhibit a modest level of star formation, such as that seen in CG 90.  
The question of where
the large population of LCBGs seen at $z =$ 0.5 to 1.0 has gone is then resolved, at least in part, by the population of less actively
star-forming galaxies seen in Figure~\ref{fig:properties}.  The large number of KISS ELGs located below the LCBG selection region
in this plot appear to be able to account for a large fraction of the ``missing" LCBGs.  The drop in density of LCBGs may be directly
linked to the fall in the star-formation rate density between $z=1$ to the present (e.g., Madau \etal 1996, 1998).  The galaxies
still exist, they are simply in a somewhat less active state.  In this scenario, LCBGs are not pathological outliers, but rather the 
most compact galaxies at the extreme end of a continuous distribution of gas-rich star-forming galaxies.  

It should be relatively simple to test the above hypothesis. A combination of high spatial resolution surface photometry in the optical
and VLA HI mapping in the radio could be used to evaluate the stellar- and gas-mass distributions of the local LCBG candidates.  A direct 
comparison of these properties between the LCBGs and a sample of ``normal" galaxies would reveal whether our LCBGs are structurally
different from the non-LCBGs.  These same observations would be extremely helpful in establishing their dynamical masses as well.
This latter parameter is essential for showing clearer linkage between the local samples of LCBGs with the intermediate- and high-redshift
analogues.  Efforts are already underway in this regard.  As a first step in determining gas and dynamical masses, we have obtained 
single-dish HI observations with the Arecibo 305-m radio telescope for the sample of LCBGs accessible from Arecibo, along with a
control sample of ``normal" star-forming galaxies.  With these data, we will derive HI masses and provide at least a rough estimate for the
dynamical masses (Werk \etal 2005) for our LCBGs.  We will also be able to ascertain which of our galaxies would be suitable for
follow-up VLA mapping.  In addition, we have begun acquiring higher resolution optical images of our LCBGs, with the goal of being
able to perform detailed surface photometry.  In addition, these more detailed images will allow us to carry out a morphological analysis 
of our sample of local LCBGs, that will provide clues as to whether their intense starbursts are caused by mergers and/or interactions, 
or are simply the result of their intrinsically compact nature.  Finally emission line-widths measured from high-resolution
spectra combined with near-infrared images and our HI mass estimates could settle the debate about the masses of LCBGs.

\section{Summary and Conclusions}

In this paper, we present a sample of local analogues to the intermediate- and high-$z$ LCBGs. We selected our sample from 
only those KISS galaxies that are actively forming stars. Due to resolution effects, we limit our sample to only those
galaxies within 185 Mpc of the Milky Way.  The sample of LCBGs is defined by the following criteria:  a surface brightness
limit of $SB_e <  21.0$ mag arcsec$^{-2}$, an absolute magnitude limit of $M_B < -18.5$, and a color cut-off of $B-V < 0.6$
with $H_0= 75$ km s$^{-1}$ Mpc$^{-1}$. A total of 16 KISS SFGs meet these criteria.

We have compared some of the properties of the KISS LCBG candidates to those of a comparison sample of KISS SFGs, and 
a comparison sample of Century Survey (CS) galaxies that are supposed to represent ``normal galaxies.''
On average, the KISS LCBG candidates have a higher tendency to emit detectable radio continuum flux, have higher H$\alpha$
luminosities indicating strong star-formation properties, have slightly lower than expected metal
abundances for KISS galaxies of their luminosity, and considerably smaller size than those galaxies in both KISS and CS
comparison samples. We have calculated the volume density for the sample of local LCBG candidates to be 
$5.4\times10^{-4}h_{75}^{3}Mpc^{-3}$.  This value indicates a substantial density evolution of LCBGs, which is in
general agreement with the predictions made by \cite{lilly96,phil97,lilly98} and \cite{cfrs}.  However, precise estimates
of the density evolution of the LCBGs are difficult, due to the large uncertainties in the values of the volume density for 
higher-redshift populations.  We have utilized the CS database in an attempt to select a sample of 
local non-active LCBGs in order to constrain the density of non-star-forming LCBGs.  We found no additional non-star-forming
LCBGs, although one weakly star-forming (possibly post-starburst) galaxy was recognized.  We postulate that LCBGs are structurally 
predisposed for star-formation, and that their compact nature may be indicative of a high central gas concentration that drives 
ongoing star-formation. 

The need for a local sample of LCBGs is well documented. Much of the published work on these galaxies speaks to the need for
a representative local comparison sample that will help settle the debates over their size and mass, morphology, and low-
redshift counterparts. We conclude that our sample of local LCBG candidates contains galaxies that are
likely to be local versions of the more distant LCBGs. This local sample provides the starting point in our ongoing study
of the properties and evolution of LCBGs.

\acknowledgments

Funding for this work was provided by an NSF Presidential Faculty Award to JJS (NSF-AST-9553020). 
Additional support for our ongoing follow-up spectroscopy campaign came from continued funding
from the NSF (NSF-AST-0071114) and Wesleyan University.   We are grateful to the anonymous referee for
several helpful suggestions that improved the presentation.  We thank the numerous KISS team members 
who have participated in spectroscopic follow-up observations during the past several years, 
particularly Caryl Gronwall, Drew Phillips, Gary Wegner, Jason Melbourne, Laura Chomiuk, 
Kerrie McKinstry, Robin Ciardullo, and Vicki Sarajedini.  Special thanks to Rafael Guzm\'{a}n for providing us 
with information about selection criteria for LCBGs, and several useful discussions.  Finally, we wish to thank 
the support staffs of Kitt Peak National Observatory, Lick Observatory, the Hobby-Eberly Telescope, MDM 
Observatory,  and Apache Point Observatory for their excellent assistance in obtaining both the survey data as 
well as the spectroscopic observations.

\clearpage

\begin{deluxetable}{ccrccccrcc}
\tablecolumns{10}
\tablewidth{0pt}
\tablenum{1}
\tablecaption{A KISS sample of LCBG candidates.} 

\tablehead{
\colhead{KISSR}&\colhead{KISSB}&\colhead{$m_{B}$}&\colhead{B-V}&\colhead{Diameter}&\colhead{$SB_{e}$}
&\colhead{Dist}&\colhead{M$_{B}$}&\colhead{log $L_{H_{\alpha}}$}&
\colhead{log P$_{1.4GHz}$}\\
\colhead{\#}&\colhead{\#}&&&\colhead{kpc}&\colhead{mag/arcsec$^{2}$}&\colhead{Mpc}
&&\colhead{ergs/s}&\colhead{Watts/Hz}\\
\colhead{(1)}&\colhead{(2)}&\colhead{(3)}&\colhead{(4)}
&\colhead{(5)}&\colhead{(6)}&\colhead{(7)}
&\colhead{(8)}&\colhead{(9)}&\colhead{(10)}
}
\startdata
113 & \nodata & 15.36 &  0.40 &   1.53  &     18.52&  93 & -19.48&  40.46    & \nodata\\
147 & \nodata & 12.44 &  0.55 &   6.53 &      20.68 & 33 & -20.18&  39.59 &  22.19\\
218 & \nodata & 15.34 &    0.49 &    4.21  &      20.93 &  84 &  -19.29&   41.20&    21.60\\
242 & 128 &  16.66 &    0.35 &    4.02&        20.84 &  153 &  -19.26 &  41.72 &   21.62\\
\nodata  & 169 &  16.56 &    0.51 &    3.39  &      20.86  & 122  & -18.87 &  41.14    & \nodata\\
707 & 189 &  14.99 &   0.40 &    5.42 &       20.85 &  94 &  -19.87 &  41.64 &   21.41\\
1094 & 218 &  15.24 &   0.41  &   6.08 &       20.60 &  134 &  -20.40 &  41.91 &   21.86\\
1271 & \nodata &  17.33 &   0.50 &    2.82  &      20.74 &  153 &  -18.60&  40.40    & \nodata\\
1274 & \nodata &  14.37  &  0.57 &    6.91 &       20.58&   99 &  -20.61 &  41.65 &   21.89\\
1400 & \nodata &  16.45 &   0.17 &    4.49  &      20.57 &  176 & -19.78 &  41.24    & \nodata\\
1438 & \nodata &  16.52 &   0.33 &    3.01 &       20.81 &  109 & -18.67 &  40.66    & \nodata\\
1578 & \nodata &  15.85 &   0.40  &   3.11 &       20.12 &  114 & -19.43 &  41.64  &  21.71\\
1637 & \nodata &  16.30 &   0.33 &    4.46 &       20.89 &  141  & -19.44 &  41.14   & \nodata\\
1664 & \nodata &  15.33 &   0.60 &    5.96 &       20.62 &  136 & -20.34 &  41.66  &  22.22\\
1870 & \nodata &  15.66 &   0.59 &    2.99 &       20.71 &  76   &-18.75 &  41.09    & \nodata\\
2079 & \nodata &  16.80 &   0.52 &    3.17 &       20.84  & 129  & -18.75 &  41.20    & \nodata\\
\enddata

\end{deluxetable}

\end{document}